\documentclass{aa}

\usepackage{epsf,epsfig}

\begin{document}

%
\title{BeppoSAX observations of soft X-ray Intermediate Polars}


\author{ D. de Martino\inst{1},
G. Matt\inst{2},
T. Belloni\inst{3},
F. Haberl\inst{4},
K. Mukai\inst{5}
}

\offprints{D.~de Martino}

\institute{
INAF--Osservatorio Astronomico di Capodimonte, Via Moiariello 16, I-80131 
Napoli, 
Italy \email{demartino@na.astro.it}
\and
Dipartimento di Fisica, Universita' degli Studi Roma Tre, Via della
Vasca Navale 84, I-00146 Roma, Italy
\email{matt@fis.uniroma3.it}
\and
INAF--Osservatorio
Astronomico di Brera, Via Bianchi 46, Merate, Italy
\email{belloni@merate.mi.astro.it}
\and
Max-Planck-Instit\"ut f\"ur Extraterrestrische Physik,
Giessenbachstra{\ss}e, Postfach 1312, 85741 Garching, Germany
\email{fwh@mpe.mpg.de}
\and
Laboratory for High Energy Astrophysics, NASA/GSFC, Code 662, Greenbelt, 
MD 20771, USA \email{mukai@milkyway.gsfc.nasa.gov}
}

\date{Received 5 August 2003; Accepted 31 October 2003}

\authorrunning{de Martino et al.}
\titlerunning{BeppoSAX observations of soft X-ray Intermediate Polars}
\markboth{...}{...}

\abstract{ 
We present broad-band (0.1--90\,keV) spectral and temporal 
properties of the three Intermediate Polars, RE\,0751$+$144\,(PQ\,Gem), 
RX\,J0558.0$+$5353\,(V405\,Aur) and RX\,J1712.6$-$2414\,(V2400\,Oph) based on 
simultaneous soft and hard X-ray  observations with the
BeppoSAX satellite. 
The analysis of their spectra over the wide
 energy range of BeppoSAX instruments allows us to identify 
the soft and hard X-ray components and to determine simultaneously their
temperatures. The black--body temperatures of the irradiated 
poles of the white dwarf atmosphere are found to
be 60--100\,eV, much higher than those found in their synchronous
analogues, the Polars. The temperature of the optically thin
post--shock plasma  is well
constrained in RX\,J1712.6$-$2414 and in RE\,0751$+$144 (13 and 17\,keV)
and less precisely determined in RX\,J0558.0$+$5353. In the first two
systems evidence of subsolar abundances is found, similarly to that
estimated in other magnetic Cataclysmic Variables. A Compton
reflection component is present in RX\,J0558.0$+$5353 and in
RE\,0751$+$144 and it is favoured in RX\,J1712.6$-$2414. Its origin is likely
at the irradiated white dwarf surface. Although these systems
share common properties (a soft X-ray component and optical polarized
radiation), their 
X-ray power spectra and light curves at different energies suggest
accretion geometries that  cannot be reconciled with a 
single and simple configuration. 
}

\maketitle 

\section{Introduction}

Intermediate Polars (IPs) are the hardest X-ray emitting Cataclysmic
Variables (CVs) containing a magnetized white dwarf (WD) that is accreting
material from a late-type Main Sequence star.  Unlike the
strongly magnetized (B$\sim$10-230\,MG) orbitally phase--locked Polars for
which $\rm P_{orb}=P_{rot,WD}$,
the magnetic field of the white dwarf is not strong enough ($\rm
B<$10\,MG) to lock its
rotation to the binary orbit and thus $\rm
P_{rot,WD} < P_{orb}$. 
IPs typically populate the long period distribution of magnetic CVs,
whilst Polars are generally found at short orbital periods. Hence,
 the wide orbits and the low magnetic fields in IPs can allow the
formation of an accretion disc, which is truncated at the magnetospheric
radius. However, other accretion configurations can occurr in
IPs, such as the complex disc-overflow, where the
stream of matter from the secondary star overpasses the accretion disc and 
is  channeled towards  the magnetic poles (Hellier 1991) or the
direct (or stream) accretion without an intervening disc. This latter
configuration is encountered in Polars, where the
accretion material directly flows towards the magnetic poles of the white
dwarf. 

\begin{table*}[ht!]     
\centering 
\caption{BeppoSAX observations of RX\,J0558, RE\,0751 and RX\,J1712}
\vspace{0.05in}
\begin{tabular}{llccccccc}
\hline
\hline
   ~ &   ~ &   ~ &     ~ &    ~ & \cr

Object & Date & MECS &  LECS & PDS & Flux$^{5}$ \cr

       &      &  $\rm T_{expo}$(ks)$^{1}$ ~~~~ Count Rates$^{2}$ &        
  $\rm T_{expo}$(ks)$^{1}$ ~~~~   Count Rates$^{2}$ &  $\rm 
T_{expo}$(ks)$^{3}$  ~~~~   Count Rates$^{4}$   \cr
\noalign {\hrule}
 ~ & ~ & ~ & ~ & & \cr
RX\,J0558 & Oct. 7-8, 1996 & 44.8 ~~~~~~~~~~~~~~~~~ 0.27 & 13.7 
~~~~~~~~~~~~~~~ 0.09 & 
84.4 ~~~~~~~~~~~~~~~~ 0.32 & 2.79 \cr

RE\,0751  & Nov. 9-12, 1996 & 114.7 ~~~~~~~~~~~~~~~ 0.32 & 27.5 
~~~~~~~~~~~~~~~ 0.15 & 110.4 ~~~~~~~~~~~~~~~~0.38 & 5.30 \cr

RX\,J1712  & Aug. 17-19, 1998 & 82.6 ~~~~~~~~~~~~~~~  ~0.50 & 33.0 
~~~~~~~~~~~~~~ 0.28 & --  ~~~~~~~ ~~~~~~~~~~~~~~~
 -- & 3.12  \cr

& ~ &    ~ &             ~ &      &   \cr
\noalign {\hrule}
\hline
\hline
\end{tabular}
~\par
\begin{flushleft}
$^{1}$: Total on-source exposure time.\par
$^{2}$: Net count rates in units of cts\,s$^{-1}$.\par
$^{3}$: PDS total exposure times relative to two units (see text).\par
$^{4}$: PDS net count rates in 13-90\,keV range.\par
$^{5}$: 0.1-10\,keV phase averaged absorbed flux in units of
10$^{-11}$\,erg\,cm$^{-2}$\,s$^{-1}$ as derived from best fit
parameters in Table\,2.\par

\end{flushleft}
\end{table*}

\noindent The X-ray emission in
IPs is generally strongly pulsed at the spin $\rm P_{rot} = 1/\omega$ 
period of the white
dwarf, indicating that material accretes onto the magnetic poles via a
disc, as it loses memory of the orbital motion (disc-fed
systems). This material flows onto the poles in an arc--shaped accretion
curtain (Rosen et al. 1988). However, not all systems are found
to pulse only  at the white dwarf rotation. The presence of 
additional frequencies in the X-ray power spectra, such as the orbital 
$\Omega$ and the beat (or
synodic) $\omega -\Omega$,  indicate
a disc-overflow accretion whose relative proportion can vary from
system to system and with time (Hellier 1991; Wynn \&
King 1992; Norton et al. 1992). Only one system, 
RX\,J1712$-$24\,(V2400\,Oph) is  known to date to
be pulsed at the beat period (Buckley et al. 1997), thus being
an unique example of a discless IP.

\noindent The X-ray emission of IPs is hard and optically thin
($\rm kT_{Brems.} \sim  5-30\,keV$), originating from the post-shock
region
above the magnetic pole. These hard
X-rays are highly absorbed by cold material (column densities up to
$\rm 10^{23}\,cm^{-2}$) within the accretion flow (Ishida et
al. 1994) and are expected to be reflected by the WD atmosphere (Beardmore
et al. 1995; Done et al. 1995; Matt 1999; Ezuka \& Ishida 1999; de Martino
et al. 2001). Unlike   Polars, 
which also show a strong soft X-ray black-body component 
(20-50\,eV), only three systems, RE\,0751$+$144\,(PQ\,Gem), 
RX\,J0558.0$+$5353\,(V405\,Aur) and the faint
RX\,J0512.2$-$3241\,(UU\,Col), were discovered as soft X-ray IPs (Mason et 
al. 1992; Haberl et 
al.  1994; Burwitz
\& Reinsch 2001). The first two are also known to show circular
polarized optical emission, suggesting a close link of soft X-ray IPs to
the
Polars. Indeed the soft X-ray emission in Polars is due to reprocessing, 
in the white dwarf atmosphere, of
hard X-rays and cyclotron radiation, which are emitted from the post-shock 
region. In particular, the relative proportion of
hard-to-soft X-ray emissions was found to depend on the magnetic field
strength and local mass accretion rate (Voelk \& Beuermann 1998). Hence, 
the soft IPs might be considered as true non-synchronous Polars and  
probably as their progenitors. Why most IPs do not show such a soft X-ray
component is still an open question.

In this paper we present BeppoSAX observations of RE\,0751$+$144\,(PQ\,Gem) 
(henceforth RE\,0751), RX\,J0558.0$+$5353\,(V405\,Aur) (henceforth
RX\,J0558)
and RX\,J1712.6$-$2414\,(V2400\,Oph) (henceforth RX\,J1712), which together
with 
RX\,J0028.8$+$5917\,(V709\,Cas) (de Martino et al. 2001) are so far
the only IPs for which a simultaneous broad-band X-ray coverage has been
performed. All three IPs are characterized by  circular
polarization in their optical emission. Also, while the first two
were discovered as soft X-ray IPs by ROSAT, RX\,J1712 was only recently 
found by XMM-Newton (Haberl 2002) to possess a soft, but hot (80-100\,eV) 
X-ray component.
Here we discuss their temporal and spectral properties over the wide
energy range of the BeppoSAX instruments.

\section{Observations and data reduction}

The BeppoSAX satellite (Boella et al. 1997) 
performed  pointed  observations of RX\,J0558 in Oct. 1996, RE\,0751 in 
Nov. 1996  and RX\,J1712 in Aug. 1998 with its co-aligned Narrow 
Field Instruments (NFI), covering
the wide 0.1-300\,keV energy range, 
as detailed in Table\,1. All three sources were
detected with the LECS (0.1-10\,keV) and MECS
(1-10\,keV) instruments. Due 
to the failure of one of the three MECS units in May 1997, RX\,J1712 MECS
data were collected by two units.  Because of LECS orbital 
limitations, 
LECS exposures were much shorter than those of the MECS (but see also 
sect.\,3.1). The 
PDS (13-300\,keV) is a collimated instrument, which monitors the
background continuously switching two (of the four) detectors with a
dwell time of 96\,s. While RE\,0751 and RX\,J0558 are detected 
up  to 90\,keV, for RX\,J1712, the presence of a 
bright  close-by cluster of galaxies,
which heavily contaminates count rates above 13\,keV, the PDS data could 
not be safely extracted.

\noindent LECS light curves and spectra  have
been extracted using a circular region centred on the sources with a 
  radius of 8\,arcmin for RE\,0751 and RX\,J1712 and 6\,arcmin
for the faint RX\,J0558. The
MECS light curves and spectra for all three sources have been 
extracted with a 4\,arcmin radius.
While the whole LECS band was used for timing analysis, the
spectra from this instrument have been analyzed only below 4\,keV due to
calibration problems at higher energies.  For both intruments, the
background was measured and subtracted using the same detector
regions during blank field pointings. 
The PDS light curves and spectra were extracted using a standard   
routine provided by the BeppoSAX Data Center (SDC). We 
conservatively extracted the data in the energy range    
13--90\,keV, above which the sources are hardly detected.
The procedure allows  rejection of particle background events,
as well as spikes caused by single-particle hits, which produce
 fluorescent cascades that are mainly recorded below
30\,keV. The background spectrum  is evaluated for each of
the two half arrays accumulating the off--source spectra. The
background light curves are constructed by linear interpolation between
two off--source pointings in each array. Then the spectra and light curves
from the two arrays are merged together to construct the
background spectrum and light curve for final subtraction.

\noindent Net on-source exposure times and count rates for 
the three sources are reported in Table\,1.

\section{Timing Analysis}

A search for periodicities was performed on light curves extracted in 
the MECS energy range between 1-10\,keV. For sake of clarity, 
the left panels of Figs.\,1,\,3 and 5 report the extracted 
light  curves with a binning time of 60\,s. They  show  
clear periodic pulses on time scale of   minutes, with 
no apparent variability on timescales of hours, implying no orbital 
modulations.

\begin{figure*}
\begin{center}
\mbox{\epsfxsize=8cm\epsfysize=8cm\epsfbox{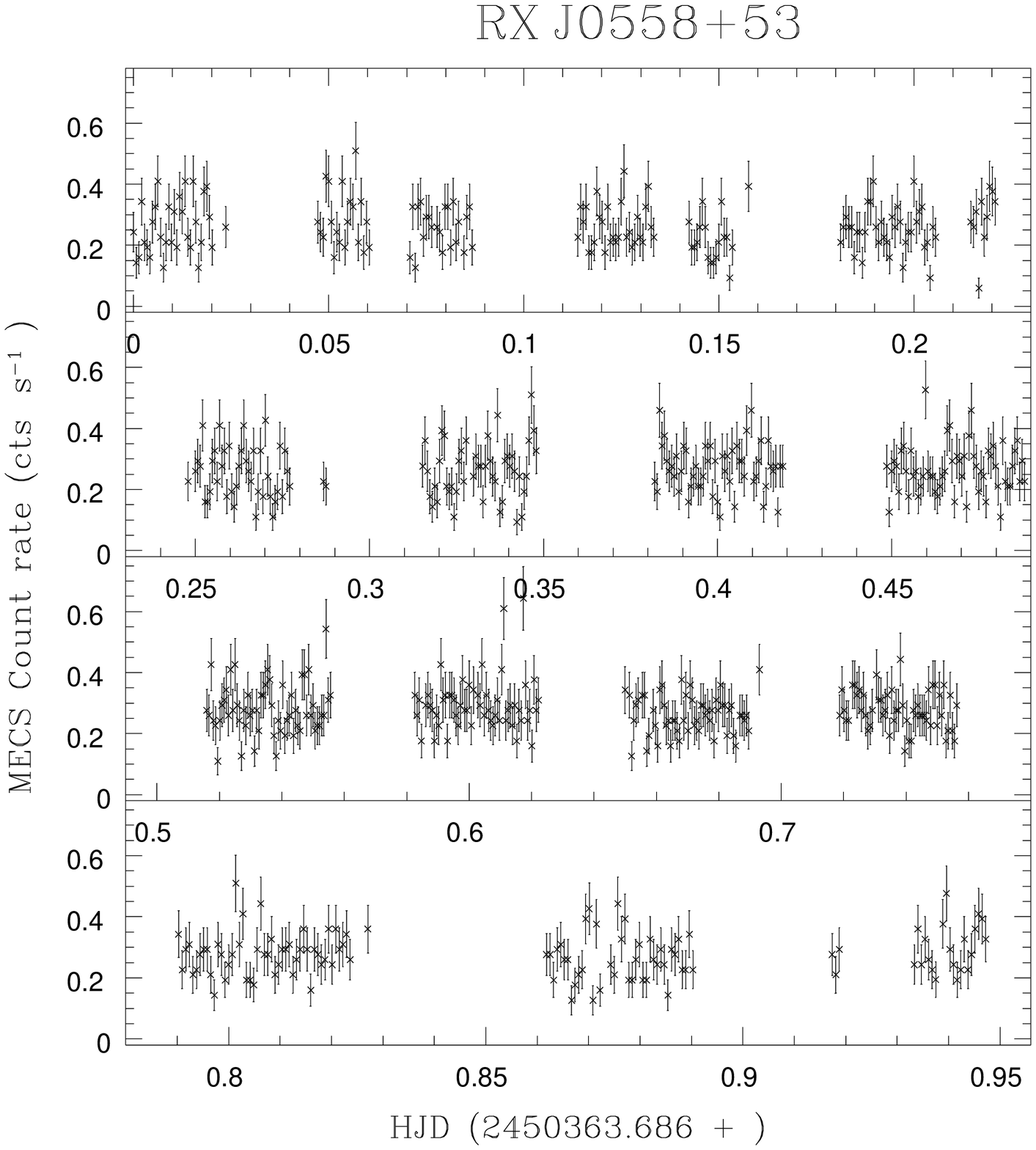}\epsfxsize=8.7cm
\epsfysize=8cm\epsfbox{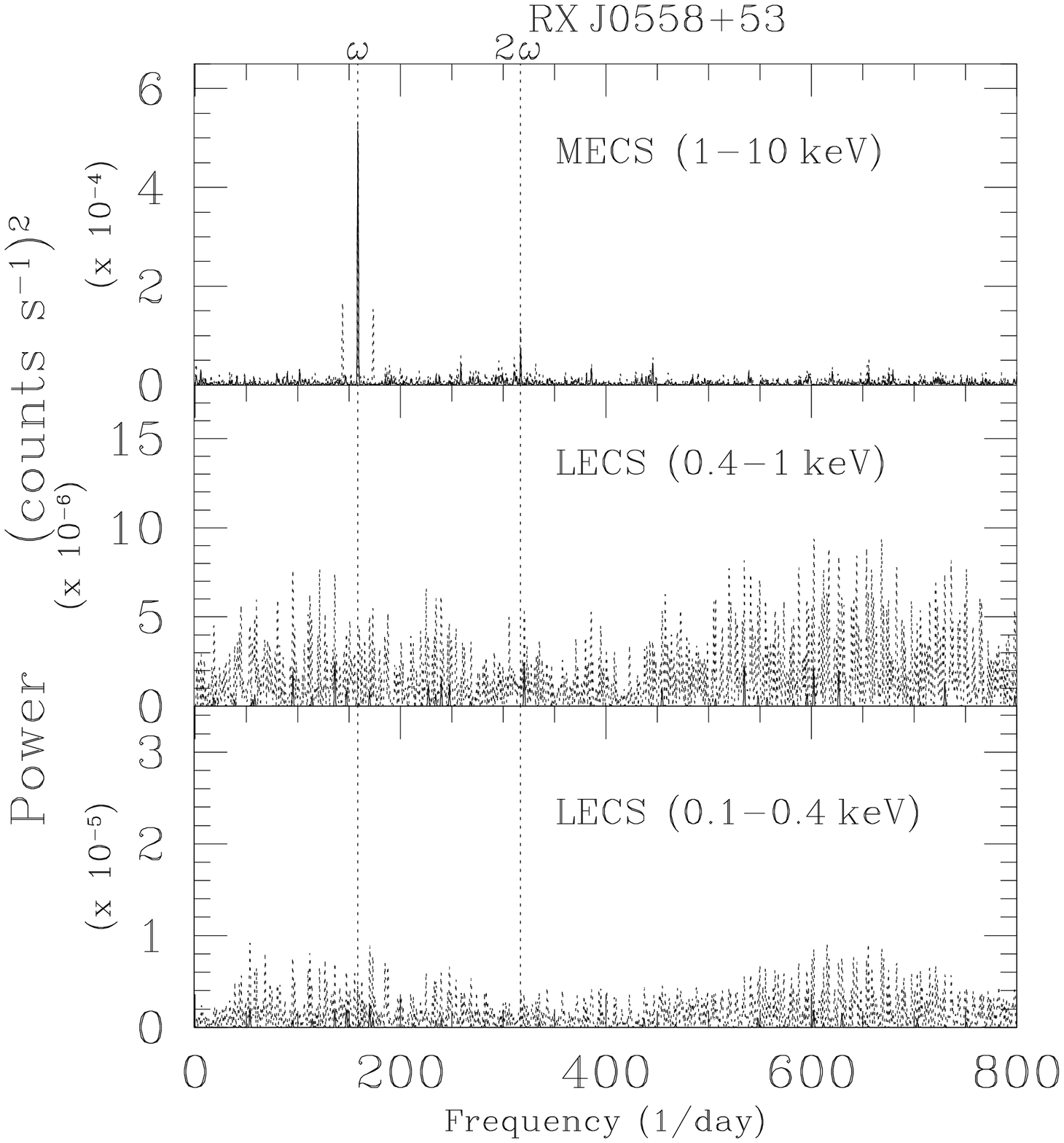}}
\caption[]{\label{rxj0558_curves} 
{\bf Left:} The net MECS 60\,s binned light
curve of RX\,J0558.  {\bf Right:} The DFT (dotted line) and CLEANED 
(solid line) spectra in the MECS (1-10\,keV), and two LECS  
(0.1-0.5\,keV) 
and (0.5-1\,keV) bands, show a strong signal 
at the 545\,s rotational period of the white dwarf only in the MECS band. 
}
\end{center}
\end{figure*}

\subsection{The spin pulse characteristics in RX\,J0558}

\noindent RX\,J0558 was discovered as a soft IP by Haberl et
al. (1994) with an X-ray period of 272\,s.
This period is also confirmed by optical
photometry (Skillman 1996, Allan et al. 1996). A 
further detailed re-analysis of the ROSAT
data showed that RX\,J0558 is 
 dominated by the 272\,s period below 0.7\,keV, whilst at higher energies 
it shows a 545\,s periodicity, 
indicating that the latter is the spin period of the white dwarf and the
soft X-ray period is related to the first harmonic of the spin frequency 
(Allan et al. 1996). Hence,   
our timing analysis was done on  LECS soft 0.1-0.4\,keV, 0.4-1\,keV 
bands and on the whole range of the MECS instrument. The light curves were 
extracted with a
binning  size of 68\,s (LECS) and 32\,s (MECS) and Fourier analysed using
the DFT  algorithm (Deeming 1975). In order to remove the windowing
effects of
unevenly sampled data due to the BeppoSAX orbit, the CLEAN algorithm 
(Roberts et al. 1987) has been  applied. The 
corresponding power spectra (the DFT and CLEANED) are shown in 
the right panels of Fig.\,1. Similarly to the ROSAT observations, they 
do not reveal the 4.15\,h orbital period. A strong peak at 
545\,s is detected in 
the MECS band, while below 1\,keV 
no periodicity is found (middle and lower panels). 
Differently from the ROSAT data,  the first harmonic is 
also clearly detected in the hard X-rays.  Consistency checks on the LECS
1-10\,keV band show that the 545\,s period is present but at a very
low level. As noted in sect. 2., the effective exposure of LECS instrument 
was very short ($\sim$ 14\,ks) over a 65\,ks of observing time, during
which no data were acquired for $\sim$16\,ks. We then performed separate
analysis of the first (19\,ks)  and second (25\,ks) interval, having an effective
on-source time of $\sim$7\,ks only. We could detect
the
545\,s modulation only in the first interval, implying
that the lack of detection of the soft X-ray modulation is due to the low
quality of LECS data and to the small number ($\sim$25) of spin cycles
covered. Here we note that ASCA observations carried out
during the same period and
simultaneously with SAX for
$\sim$5\,hrs confirm the presence of the 272\,s modulation in the soft
0.4-0.6\,keV band  with an amplitude of $\sim 34\%$ 
 (Mukai et al. 1997). Hence the LECS soft
X-ray band for this source will not be
discussed further for variability purposes.

\begin{figure*}
\begin{center}
\mbox{\epsfxsize=8cm\epsfysize=8cm\epsfbox{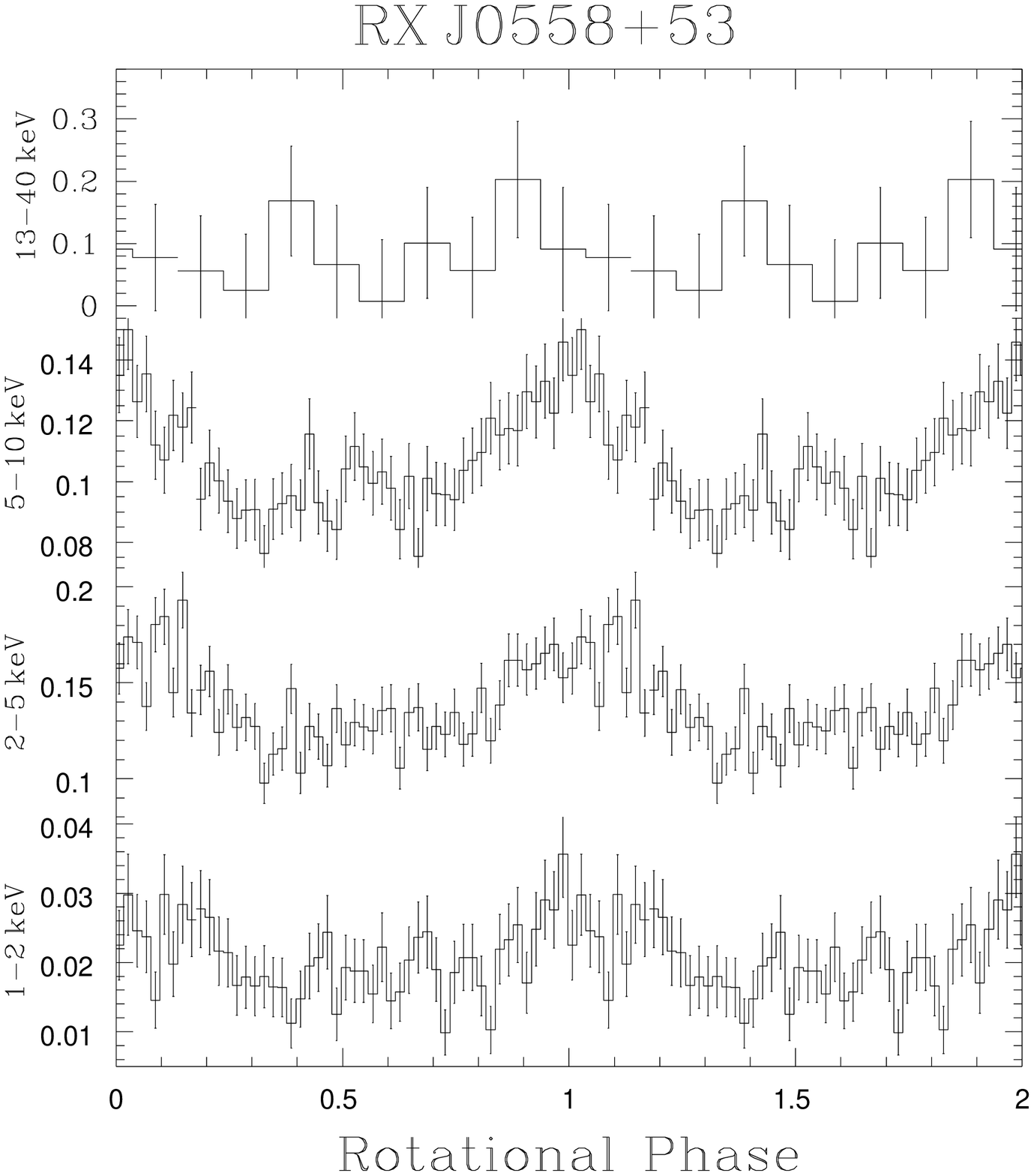}
\epsfxsize=8.7cm\epsfysize=8cm\epsfbox{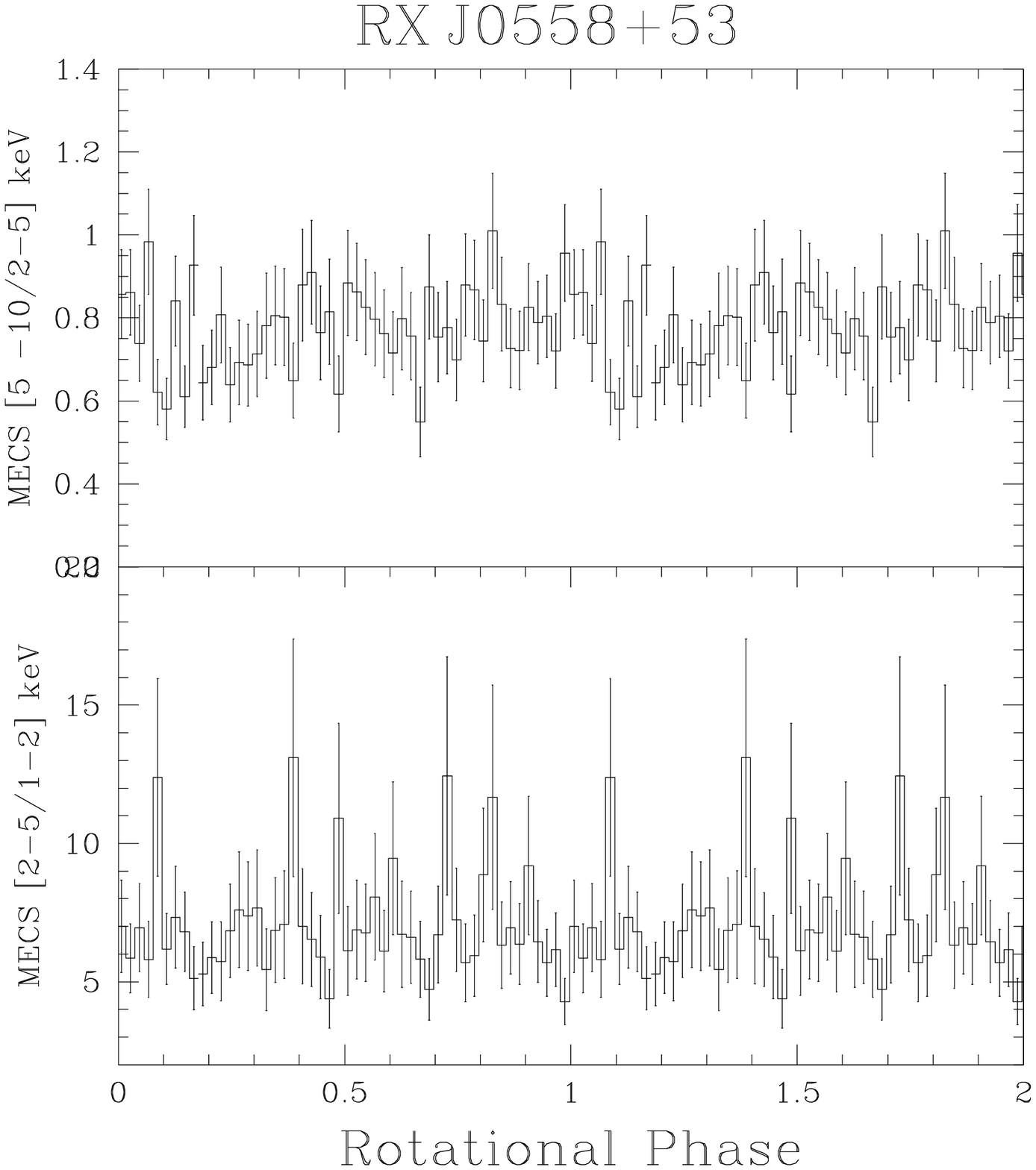}}
\caption[]{\label{rxj0558_folded} 
{\bf Left:} Folded light curves for RX\,J0558 in different energy ranges 
from 1 to  40\,keV using the ephemeris quoted in the text. {\bf Right:}  
The   hardness ratios in MECS bands show no spectral dependence 
of the spin pulsation (see text).
}
\end{center}
\end{figure*}

\noindent Allan et al. (1996) provide a spin ephemeris based on the soft X-ray
double-humped light curve. The time of their maximum, however, does 
not correspond to the maximum observed in the hard X-rays and therefore
the
 MECS light curve has been  fit with a composite sinusoidal function
accounting for the base and second harmonic frequencies. We derive:

\begin{center}
$\rm HJD_{max} =  2450364.159695(2) + 0.006311(6)xE$.
\end{center}

\noindent This ephemeris  is less precise
 than the one derived from 
ROSAT
data (Allan et al. 1996), probably due to the low statistics of the 
BeppoSAX  light  curve and to the non sinusoidal shape of the hard X-ray 
pulsation. However, it
 allows a proper phasing of the hard X-ray maximum as
shown in Fig.\,2, left panel. As for the LECS instrument,
the lack of modulation in the high energy PDS data  is probably due 
to  the limited quality. As expected, the modulation in the
1-10\,keV
band is not sinusoidal with a broad shoulder maximum and
a shallow minimum and hints of a  secondary pulse feature. The pulsed
fraction (max-min/average) is 56$\%$ in the 1-2\,keV band and $\sim
40-45\%$ in the 
2-5\,keV and 5-10\,keV bands. The hardness ratios
[2--5/1--2\,keV] and [5--10/2--5\,keV],
tracing the  hard component (right panels of Fig.\,2)  do not show
significant 
variations with spin phase, confirming the lack of energy-dependent 
modulation in the hard bands. The spin variability of the hard component
is then due to aspect angle changes of the emitting region.

\subsection{The spin modulation in RE\,0751}

RE\,0751 was the first soft X-ray IP to be discovered from the ROSAT
survey (Mason et al. 1992).  A period derivative $\rm \dot
P=1.1\times10^{-10}\,\,s\,s^{-1}$ was subsequently found from X-ray and
optical timings of the 833\,s white dwarf rotational period spanning 
 five years (Mason 1997),  indicating that the white dwarf is
spinning down at the largest
rate ever found in an IP. Its behaviour in both soft and 
hard X-ray bands has never been monitored simultaneously.  The BeppoSAX 
MECS (1-10\,keV ) light  curve reported in Fig.\,3 (left panel) clearly 
shows the  presence of rapid pulses and the absence of variability at the
5.2\,h orbital period. The DFT and CLEANED power spectrum
in the LECS  0.1-0.4\,keV, 0.4-1\,keV and MECS 1-10\,keV range are 
reported in the right panels of Fig.\,3. 
Above 1\,keV, the power spectrum shows a strong signal at the 833\,s
spin period and higher harmonics up to the third. In the 0.4--1\,keV 
range, instead, the signal at these frequencies is at the noise level.
 Here the CLEAN algorithm does not detect even the base frequency.  
Below 0.4\,keV  the signal at the base  
frequency is strong again, but not those at the higher harmonics. This
indicates 
that the 833\,s  modulation
is structured in the hard X-rays, then mostly disappears at decreasing
energies and appears again as a strong single pulse below 0.4\,keV.

\begin{figure*}
\begin{center}
\mbox{\epsfxsize=8cm\epsfysize=8cm\epsfbox{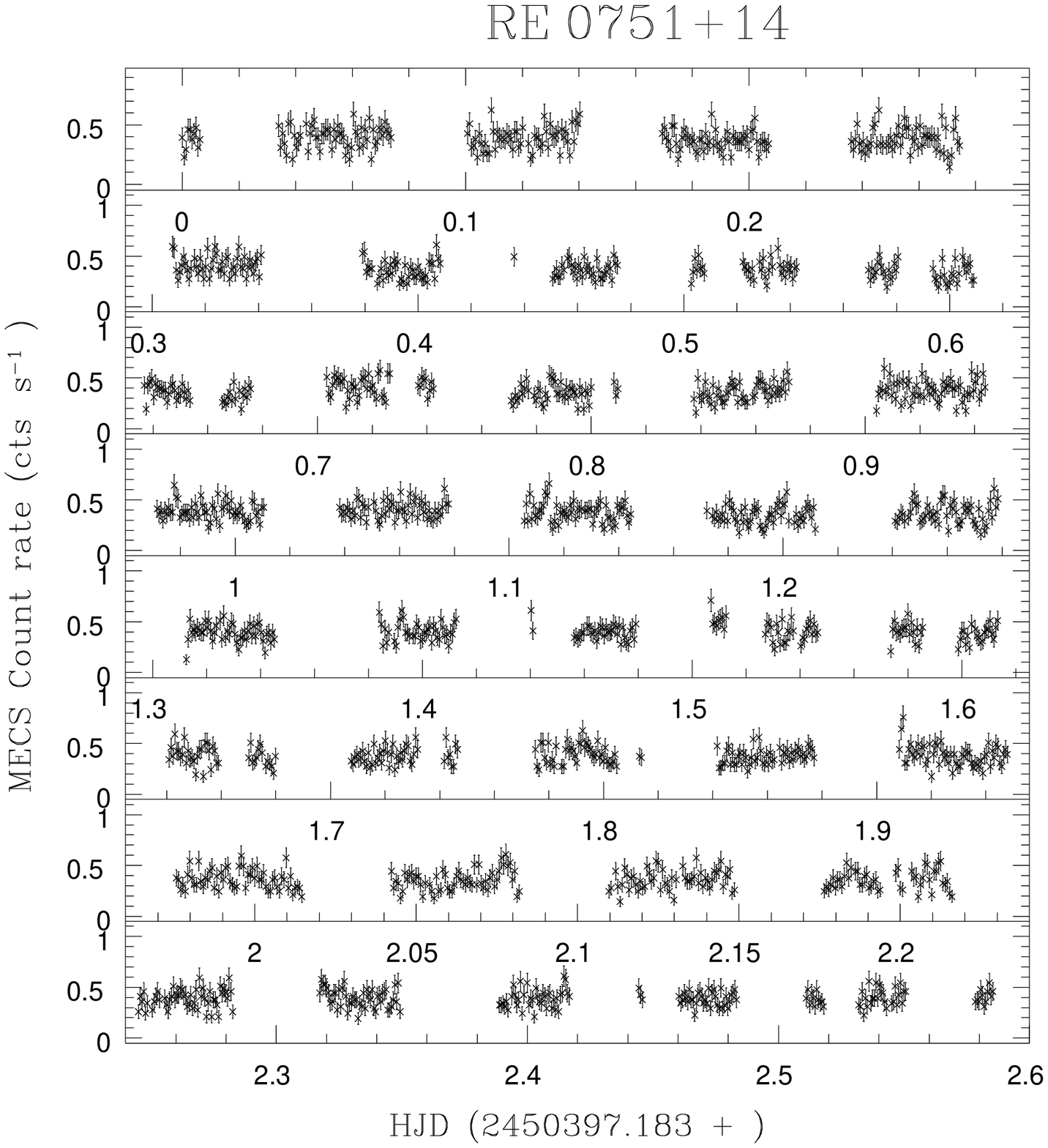}\epsfxsize=8.7cm
\epsfysize=8cm\epsfbox{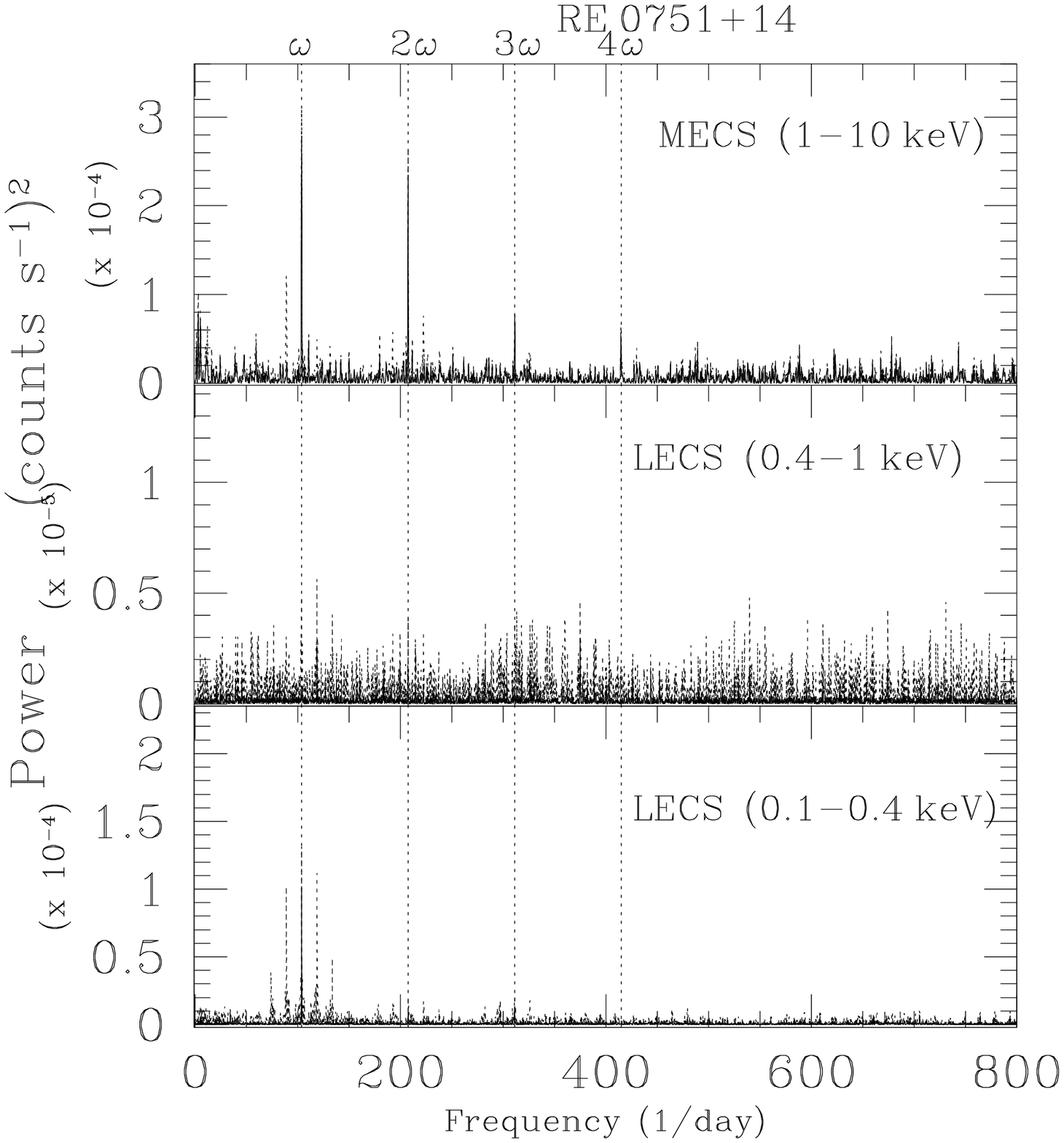}}
\caption[]{\label{pqgem_curves} 
{\bf Left:} The net  60\,s binned light
curve of RE\,0751.  {\bf Right:} The DFT (dotted line) and CLEANED 
(solid line) spectra in the MECS (1-10\,keV), and two LECS  
(0.1-0.4\,keV) 
and (0.4-1\,keV) bands. The hard X-rays show strong signals at the 833\,s 
frequency and higher harmonics. The softer band is instead dominated by 
the spin frequency only, while in the 0.4-1\,keV band the spin signal is 
not significant.
}
\end{center}
\end{figure*}

\begin{figure*}
\begin{center}
\mbox{\epsfxsize=8cm\epsfysize=8cm\epsfbox{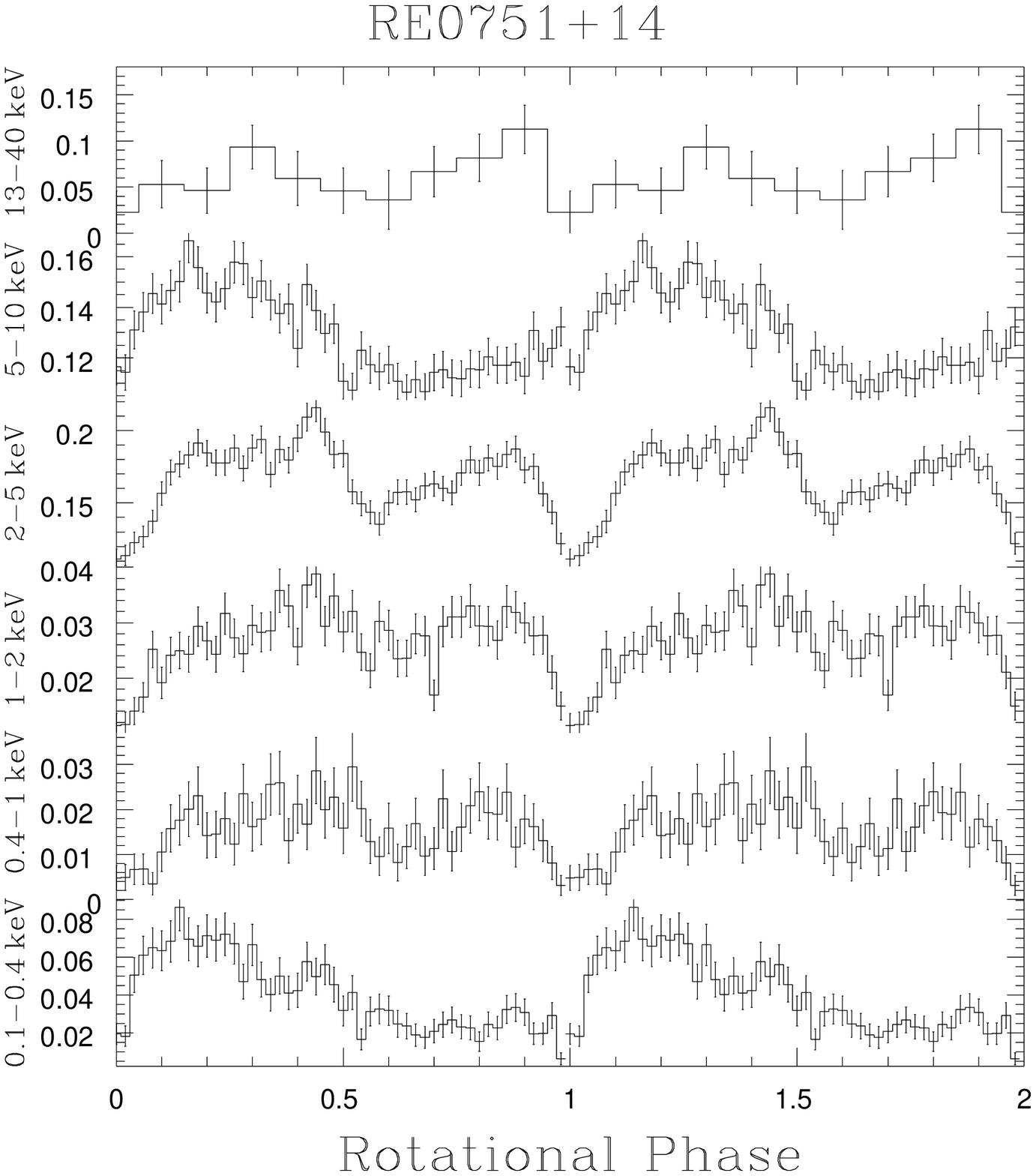}
\epsfxsize=8.7cm\epsfysize=8cm\epsfbox{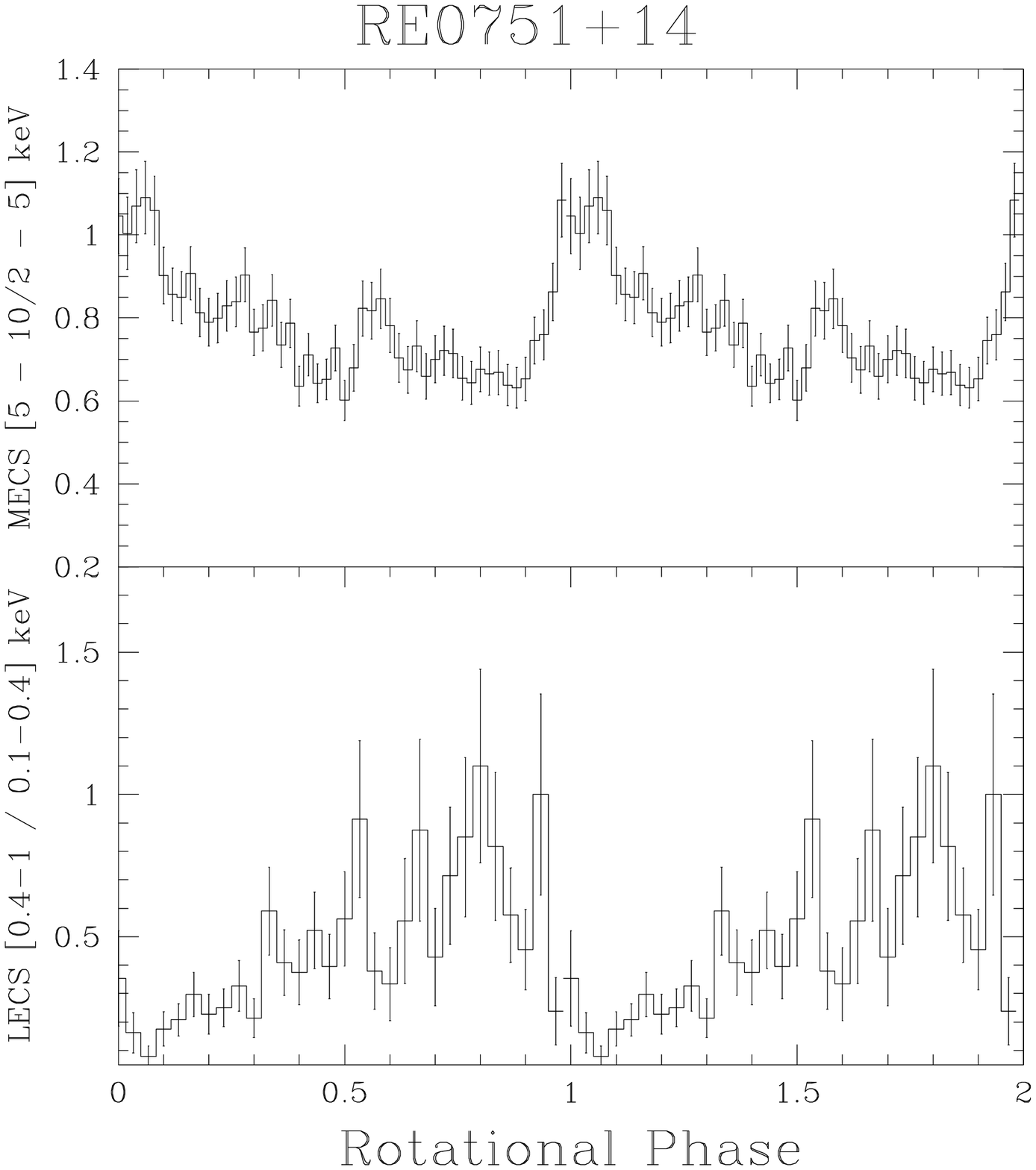}}
\caption[]{\label{pqgem_folded} 
{\bf Left:} Folded light curves of RE\,0751 in different energy ranges 
from 0.1 to  40\,keV using Mason's quadratic ephemeris (1997). {\bf 
Right:}  
The   hardness ratios in two soft LECS bands and two 
MECS bands show substantial  differences.
}
\end{center}
\end{figure*}

\noindent The energy dependent behaviour of the spin modulation has been 
inspected 
in different energy bands from 0.1 to 40\,keV. In Fig.\,4, we report
the folded rotational pulses using the quadratic ephemeris provided by 
Mason (1997), which  is based on the stable presence of a narrow dip
and is accurate enough to extrapolate to the BeppoSAX observation epoch
(i.e.
the propagated phase error is $\delta \phi$=0.03).
The light curves confirm the very complex morphology, detected
previously by  ROSAT, Ginga,  ASCA and RXTE (Duck et al. 1994; Mason 
1997; Kiziloglu et al. 1998; James et al. 2002). In the
  0.1-0.4\,keV band (as well as in the 0.1-0.5\,keV one, used by 
Duck et al. 1994) the pulse profile
is single peaked, with count rates increasing by a factor of 4, with a
hardly detectable narrow dip of  
0.06$\pm$0.01 width in phase.  In the higher 0.4-1\,keV range, the dip
increases in breadth (0.1 in phase) and in place of the broad minimum 
observed in the softer band, a secondary maximum is observed, making
the overall modulation highly structured, with large excursions throughout 
the spin cycle.  
 The modulation at energies higher than 1\,keV  shows an   
increase  in depth of the dip feature up to 5\,keV. When compared to
GINGA or ASCA data, the pulse is not double-peaked with 
primary and secondary maxima of the same intensity, but rather
consists of a more structured primary maximum and a weaker secondary
maximum which is detected up to 5\,keV. Alternatively, a
secondary
dip at $\phi$=0.55, more pronounced in the 2-5\,keV range, might
be present, simulating
a double-peaked light curve.
 The modulation becomes very sinusoidal at 5-10\,keV, 
with no apparent hint of a secondary maximum, which is instead observed by 
Ginga and broadly resembles that observed by RXTE (James et al. 2002). In
the harder band (10-40\,keV), again, we cannot detect any modulation 
due  to the limited statistics of PDS data.  The 
BeppoSAX data confirm the energy dependence of the dip which is 
not visible above 5\,keV.
 The hardness ratios [5--10/2--5\,keV] and
[0.4--1/0.1--0.4\,keV] show a different behaviour.  Above 2\,keV,
the spectrum hardens 
during the dip and it  is still hard at spin maximum and is softer at
spin minimum. A hardening is also observed around phase 0.55. 
This confirms that the dip as well as the underlying 
modulation is energy dependent.  There is probably an additional
absorption component that produces the secondary dip, thus indicating that
the absorbing
material in the accretion flow is very structured. On the other hand, 
the soft X-rays harden at spin minimum, 
implying that either the X--ray reprocessing region is hardest, 
or that there is an
additional contribution of a harder emitting region seen at spin
phases between 0.6--0.9. This aspect will be reconsidered in sects.\,4 and
5.

\begin{figure*}
\begin{center}
\mbox{\epsfxsize=8cm\epsfysize=8cm\epsfbox{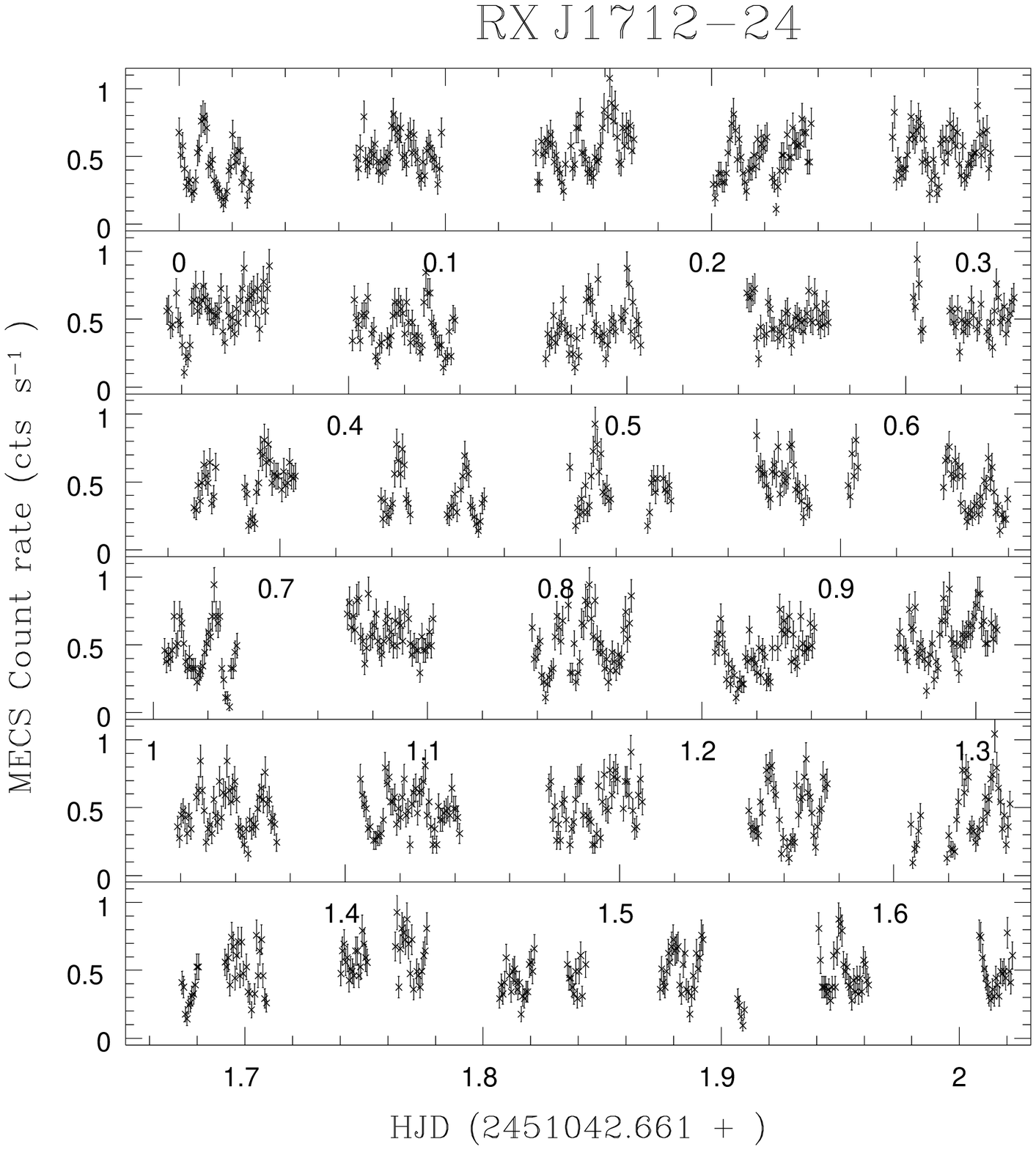}\epsfxsize=8.7cm
\epsfysize=8cm\epsfbox{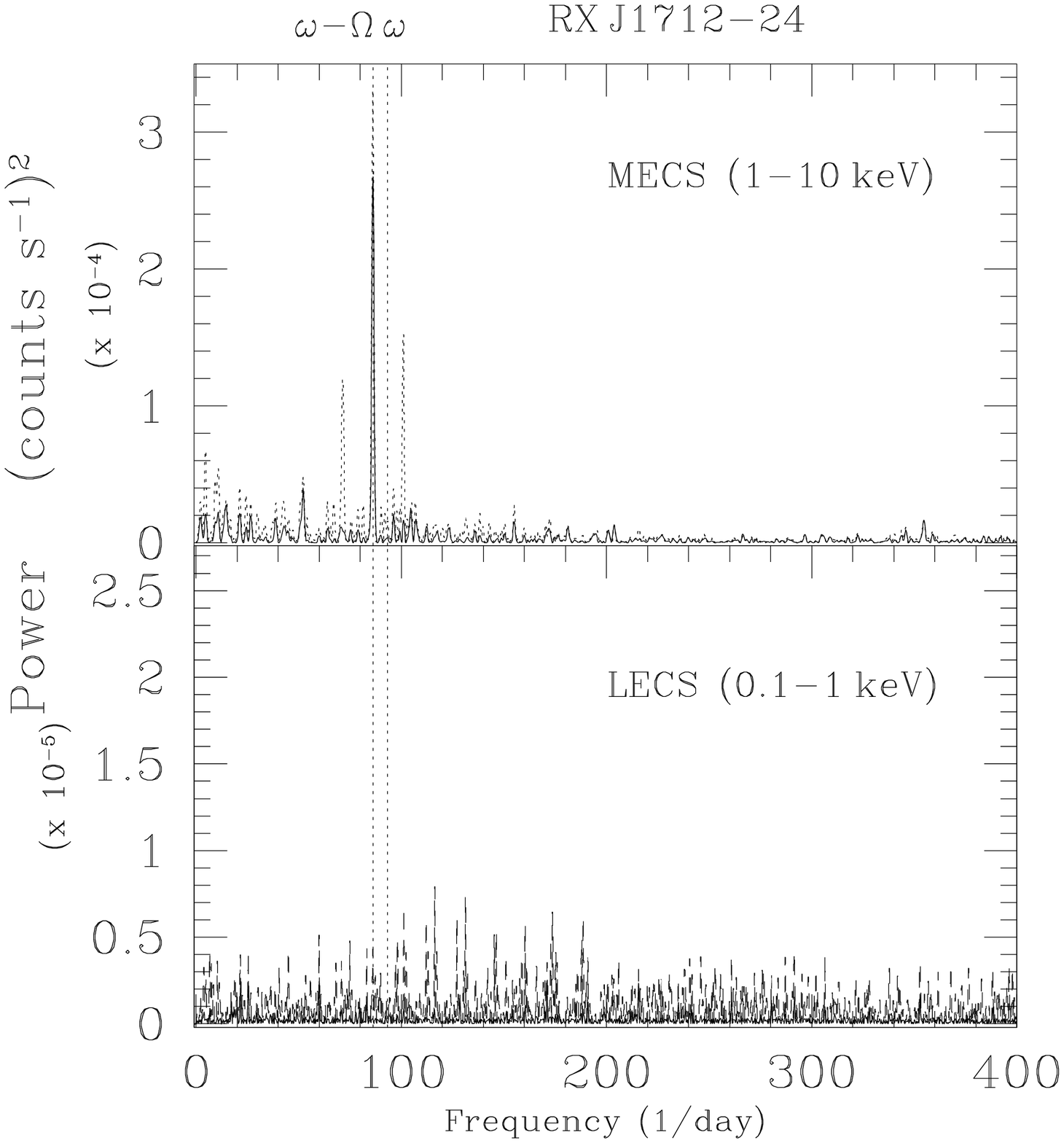}}
\caption[]{\label{rx1712_curves} 
{\bf Left:} The net MECS  60\,s binned light
curve of RX\,J1712.  {\bf Right:} The DFT (dotted line) and CLEANED 
(solid line) spectra in the MECS (1-10\,keV), and LECS  
(0.1-1\,keV)  bands. The hard X-rays are strongly modulated at the 1003\,s
synodic period while the soft X-ray band appears to be unmodulated at 
this frequency. 
}
\end{center}
\end{figure*}

\subsection{The beat pulsation of RX\,J1712}

This system was discovered as a hard X-ray IP (Haberl \& Motch 1995) 
pulsating at a period of 1003\,s, but with a 927\,s period in its 
optical polarised light   
(Buckley et al. 1995; 1997). This led to the identification of the longer
1003\,s 
period as the  synodic period  between the 3.42\,h orbital and the 927\,s 
white dwarf spin periods, thus making RX\,J1712 the only discless IP 
known so far, where accretion occurs via pole-flipping each half of a beat
cycle. 
The MECS (1-10\,keV) band light curve is shown in the left panel 
of Fig.\,5.  The DFT and CLEANED power 
spectra are reported in the right panels of that figure. A strong signal
at 
the synodic frequency is detected in the harder bands ($>$1\,keV), whilst
in the softer 0.1-1\.keV range this  frequency is  not 
detected, although evidence of power is found between between 100 and 
190\,$\rm d^{-1}$. No power is observed at the 927\,s spin and 3.42\,h 
orbital frequencies, confirming the ROSAT and RXTE results (Hellier \&
Beardmore  2002). Consistency checks on the LECS data have been carried
out and the strong beat pulsation is present in different bands above
1\,keV. 

\begin{figure*}
\begin{center}
\mbox{\epsfxsize=8cm\epsfysize=8cm\epsfbox{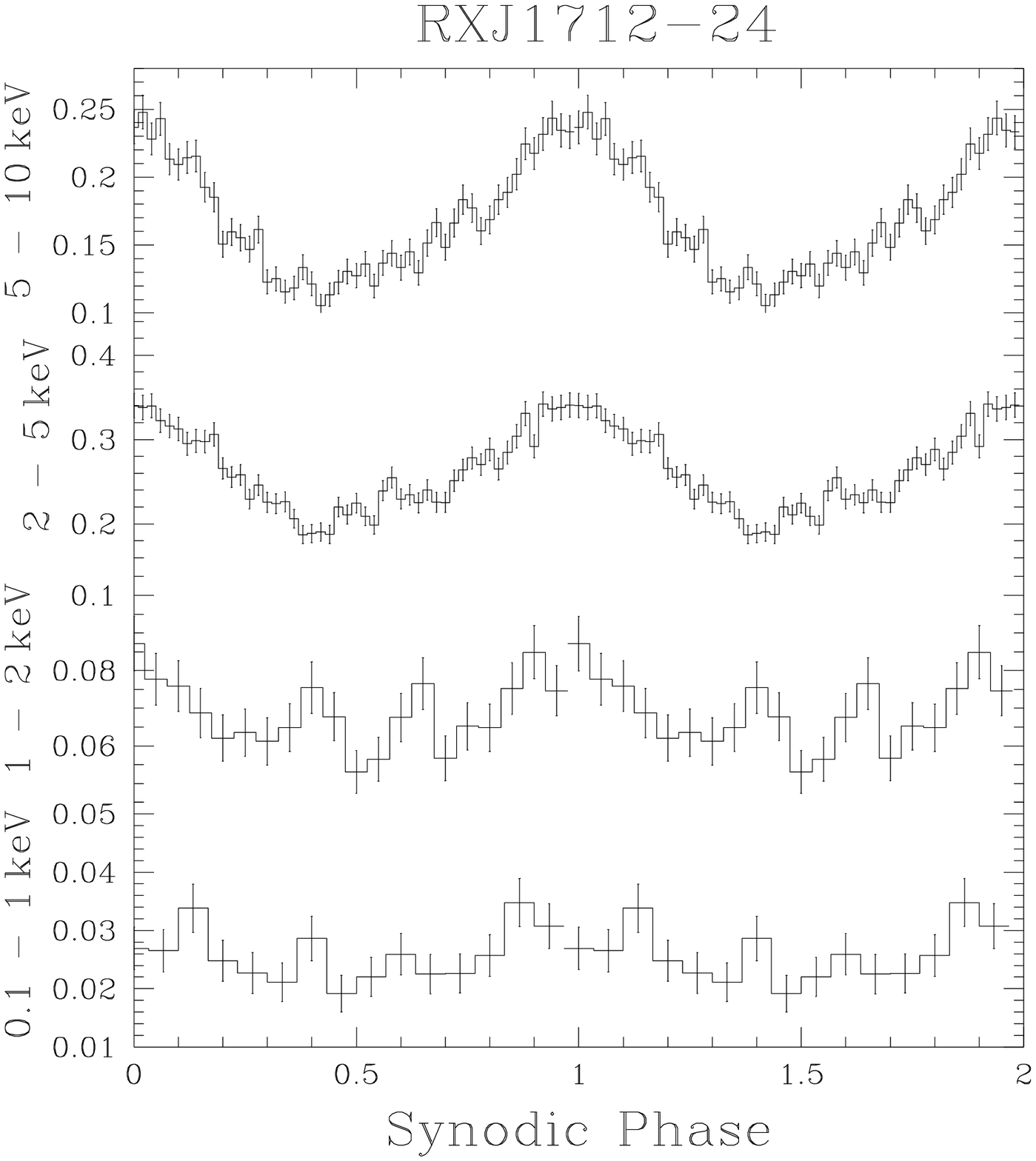}
\epsfxsize=8.7cm\epsfysize=8cm\epsfbox{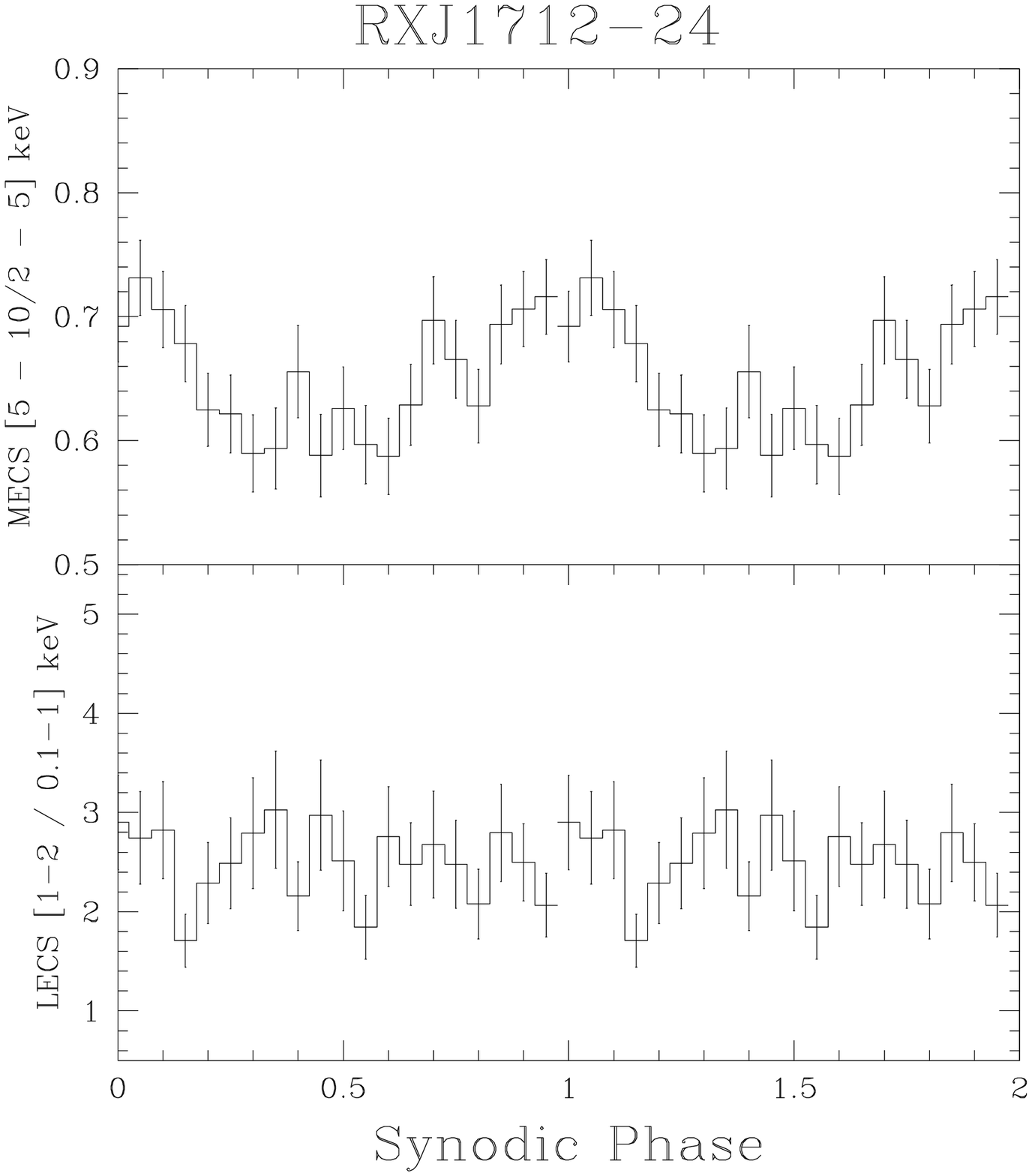}}
\caption[]{\label{pqgem_folded} 
{\bf Left:} Folded light curves of RX\,J1712 in different energy ranges 
from 0.1 to  10\,keV. {\bf 
Right:}  
The   hardness ratios in two soft LECS bands and two 
MECS hard bands showing different behaviour with respect to other IPs.
}
\end{center}
\end{figure*}

\noindent To study the energy dependence of beat pulses, 
we obtained the time of X-ray maximum for our BeppoSAX data set by 
a sinusoidal fit to the MECS (1-10\,keV) light curve: 
$\rm HJD_{max}$=2451063.62405(5). 
We then folded the light curves in different energy
bands as shown in Fig.\,6 (left panel). The modulation is stronger in the 
hard bands, with  a 
pulsed fraction of 80$\%$ in the 
5-10\,keV range, 60$\%$ in the 2-5\,keV band and 50$\%$ in the 1-2\,keV 
range. In this latter range the modulation is not 
sinusoidal, with a structured minimum. The 
0.1-1\,keV range instead shows excursions up to 60$\%$ over the
synodic cycle, although  an underlying sinusoidal trend might be
present.
The softer bands (below 2\,keV) might be indicative of  
a different emission component. 
A hardening at pulse maximum is observed above 2\,keV, while at lower 
energies no variations are detected (Fig.\,6, right panels). 
The hardening at pulse maximum in this discless system is similar to 
that  observed in Polars (Matt et al. 2000), which accrete via a
column fed directly from the stream. 
Hence, pulse maxima are observed when the 
column points towards the observer and the absorption is 
maximum, with  the hardest regions  contributing most. The lack of
strong pulsations below 1\,keV  and the fact that the X-ray modulation 
does not vanish at pulse minimum (which is seen 
in Polars) suggest an additional source of X-rays. This aspect will 
be discussed in sect. 5.

\begin{table}[h!]     
\caption{Spectral fits to the grand--average spectra of 
RE\,0751, RX\,J0558 and RX\,J1712. The adopted model is described in the 
text. Quoted errors refer to 90$\%$ confidence level for
 the parameter of  interest.}
\begin{tabular}{cccc}
\hline
\hline
& & &\cr  
Parameter & RE\,0751+14  & RX\,J0558+53  & RX\,J1712-24 \cr
& & &\cr  
\hline
& & &\cr  
$\rm N_H^{1}$ & 0.58$^{+1.27}_{-0.56}$ & 3.7$^{+2.6}_{-2.1}$ & 
$46^{+12}_{-13}$ \cr
& & &\cr  
$\rm N_{H(pcfabs)}^{2}$ &  4.8$^{+2.8}_{-1.7}$ &  
5.1$^{+2.7}_{-2.3}$ & 11.8$^{+4.1}_{-3.3}$ \cr
& & &\cr  
 $\rm C_{F}$ & 0.39$^{+0.09}_{-0.08}$ & 0.42$^{+0.05}_{-0.04}$ & 
0.34$^{+0.05}_{-0.06}$ \cr
& & &\cr  
$\rm kT_{BB^3}$ & $56^{+12}_{-14}$   & $73^{+14}_{-14}$ &  
$103^{+11}_{-9}$ \cr
& & &\cr  
$\rm kT_{hard}^4$ & 17$^{+2}_{-2}$  & 
34$^{+21}_{-11}$ & 13$^{+1}_{-2}$ \cr
& & &\cr  
 $\rm L_{BB}/L_{hard}$ & $\sim$0.4 & $\sim$0.8 & $\sim$0.7\cr
& & &\cr  
 A & 0.33$^{+0.07}_{-0.09}$ & 1.17$^{+0.93}_{-0.55}$  
& 0.36$^{+0.12}_{-0.10}$  \cr
& & &\cr  
 EW$^5$  &  188$\pm56$ & 260$^{+72}_{-80}$ & 
157$^{+37}_{-34}$ \cr
& & &\cr  
R$^6$ & 2.2$^{+1.1}_{-1.1}$ & 1.2$^{+1.7}_{-1.0}$ & 1. \cr
& & &\cr  
\hline
& & &\cr
$\chi^2_{red}$  & 1.07 & 1.22 & 1.01 \cr
& & &\cr  
\hline
\hline
\end{tabular}
\par
\begin{flushleft}
$^1$ Hydrogen column density accounting for the interstellar absorption in
units of 10$^{20}\,\rm cm^{-2}$.\par 
$^2$ Column density of the partial absorber in units of 
10$^{22}\,\rm cm^{-2}$.\par
$^3$ Temperature in units of eV.\par
$^4$ Temperature in units of keV.\par
$^5$ Equivalent width of the 6.4\,keV Gaussian in units of eV.\par
$^{6}$ Relative normalization of the reflection component in units of 
2$\pi$. It is fixed in RX\,J1712 (see text).
\end{flushleft}
\end{table}

\section{The broad-band spectral properties}

We studied the broad-band spectra  by means
of spectral fits (using XSPEC) to the phase--averaged spectra.
We  used the combined LECS, MECS and PDS data for RX\,J0558 and 
RE\,0751 allowing the study from 
0.1 to 90\,keV. For RX\,J1712 we used the LECS and MECS data only and 
hence spectra are limited from 0.1 to 10\,keV. 
 We have adopted a composite model 
consisting 
of an isothermal plasma emission (MEKAL)  with temperature $\rm 
kT_{hard}$ and metal abundance A (in number with respect to 
solar), a black--body component with temperature $\rm kT_{BB}$, two cold
media consisting of a simple absorber with column density $\rm
N_{H}$ and a partial 
covering absorber with column density $\rm N_{H(pcfabs)}$ and covering
fraction  $\rm C_{F}$. The model also includes  
a Gaussian fixed at E=6.4\,keV  to account for the iron K$_{\alpha}$ 
fluorescent line and a Compton reflection 
component, with relative 
normalization R, which  represents the solid 
angle subtended by the material in units of $2\,\pi$ for an ave\-ra\-ge 
viewing angle fixed to 60$^{o}$. The iron and other element abundances 
of this component (REFLECT in XSPEC) have been linked to the abundance of 
the MEKAL model. A Compton 
reflection   component has been demonstrated to be present in mCVs
(both Polars
and IPs) (Done et al. 1995; 
Matt 1999; Ezuka \& Ishida 1999; Matt et al. 2000; de Martino et
al. 2001). Here we note that 
the BeppoSAX data are not of enough quality to derive the 
multi--temperature structure of the post--shock region. Indeed  a power 
law temperature profile (CEMEKL in XSPEC) gives worse fits with respect
to an isothermal plasma. Hence, we are unable to estimate the
white dwarf masses as we lack  reliable shock
 temperatures.   
In  Table\,2, 
we report the results of the spectral fits with the mentioned model,
which are shown in Fig.\,7.
Here we note that for RX\,J1712, because of the lack of PDS data, we first
assumed no reflection (i.e. R=0). We also allowed for this parameter to
change and obtained an upper limit of 0.16. This is however unrealistic
for the large EW of the iron line and the low abundances and, therefore,
we assumed R=1.

\begin{figure}
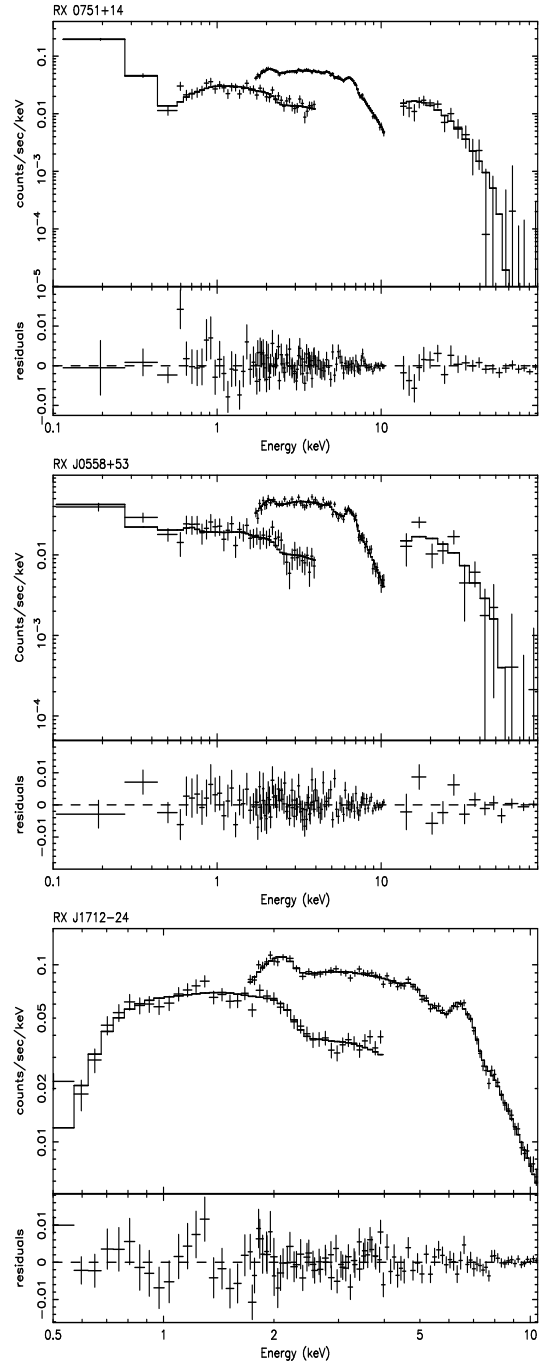

\begin{center}
\epsfig{file=0160_f7a.ps, height=7.cm, width=6.cm, angle=-90}
\epsfig{file=0160_f7b.ps, height=7.cm, width=6.cm,angle=-90}
\epsfig{file=0160_f7c.ps, height=7.cm, width=6.cm,angle=-90}
\caption[]{\label{spectra} 
{\bf From top to bottom:}  The phase--averaged spectra of RE\,0751,
RX\,J0558 and RX\,J1712 along their best fit model described in Table\,2.
Lower panels show the residuals. 
}
\end{center}
\end{figure}


\noindent For all three sources, the column densities of the simple
absorber are compatible within errors with the ROSAT determinations 
(Haberl \& Motch 1995; Duck et al. 1996) and are also in
agreement with the estimates of $\rm E_{B-V}$ obtained from UV spectra
(Mouchet et al. 1998; de Martino et al. 1998). These are lower or at most
equal to the galactic column densities in the direction of the three
sources, implying that the simple absorber can be regarded as of
interstellar origin. From the fits, 
the partial covering (up to  40$\%$) cold absorber has
instead   much higher densities than that of the galactic medium and its
origin is intrisic to the systems. As will be seen in sect.\,4.1, this
 intrinsic absorber is likely to be more complex, but this
approximation allows us to identify in a simple way, the
presence of localized cold material close to the X-ray source. 

\noindent The 
black--body component is found to be ubiquitous in all three systems with 
temperature that are at the higher end of the temperature range found in
Polars (20--60\,eV). The hottest
black--body component is found in RX\,J1712 (100\,eV) which confirms the
surprising results of Haberl (2002) obtained from XMM--Newton observations
for this source and 1RXS\,J154814.5-452845  (Haberl et
al. 2002).  Very hot temperatures are
not 
observed in Polars.  The relatively high black--body temperature in
RX\,J0558 is within errors compatible with the earlier ROSAT results
by Haberl $\&$ Motch (1995).  For RE\,0751, our spectral fit gives a
slightly higher temperature but is still consistent within errors with
that
derived by Duck et al. (1994) from ROSAT observations at 46\,eV, 
but not with the one determined from ASCA (James et al. 2002) which
however  was not well
constrained.
 Here we note that James et
al. (2002) also fix the temperature of this component to the ROSAT
value in their ASCA phase--average spectral fit, but in the
phase--resolved analysis the temperature is fixed to the value of
83\,eV, found for the spectrum at maximum. The
normalizations of the black--body components are loosely constrained
(60-70$\%$ uncertainty).

\noindent The average temperature of 
the optically thin post--shock plasma is well constrained except for 
RX\,J0558 which also has a poorly determined abundance. For this parameter 
we also find indication of underabundances in RX\,J1712 and RE\,0751.
Ezuka \& Ishida (1999) also find that magnetic CVs generally display
subsolar iron abundances.  
The temperatures of the hard X-ray component are substantially lower than
those derived excluding the reflection, which is however required by the
data.
The 6.4 keV iron line EWs are too large to be accounted for by the partial
covering absorber (Inoue 1985; Matt 2002), and they are likely produced by
the
white dwarf surface along with the Compton reflection component (Matt
1999),
as also found for the IP RX\,J0028.8+5917 (de Martino et al. 2001). For
RE\,0751,
the line EW is consistent with the measured value of R, even allowing
for the low iron
abundance (Matt et al. 1997). The same is true for RX\,J0558 when the
rather large
errors are considered, even if the value of EW, taken at face value,
appears rather
large when compared to R.
For RX\,J1712, the value of R is very loosely constrained
due to the lack of usable PDS data; for this source, a non-negligible
contribution from
the partial covering absorber is also expected, given the large column
density.
Although poorly constrained, the ratios between bolometric black--body and
hard X-ray components are
found to be close to unity in RX\,J0558 and RX\,J1712, and  lower in 
RE\,0751.

\subsection{Phase--resolved spectral variations}

To investigate  in more detail the spectral behaviour of 
pulsations, we  performed  fits 
to the spectra at maximum and minimum of  the pulse cycle of the three
sources
and also to the 
dip spectrum of RE\,0751. We used only LECS and MECS data due 
to the lack of variability in the PDS. The spectral fits consist of 
a composite model of a MEKAL plus black--body component, the 
interstellar column density, the partial absorber and a Gaussian line 
fixed at 6.4\,keV. For RE\,0751 and RX\,J0558, we also included the 
reflection component as this is expected to be angle--dependent (Matt 
1999). We have therefore fixed the interstellar hydrogen column 
densities to the values found from the fits of the phase--averaged 
spectra. Also the 
temperature and abundance of the optically thin plasma are  fixed at the 
values obtained from the phase--averaged spectra. This was done because
the  former is  an average temperature over the emitting post--shock 
region and the latter is not expected to change with phase.  
The  temperature of the black--body,  the 
partial covering parameters, the normalization of the 
Gaussian and those of the hard and black--body components as well as the 
reflection parameter  are kept free. In 
Table\,3 we report the results of the spectral 
fits for the three systems.\\

\noindent RX\,J1712 shows an increase of the partial covering absorber
(both in column density and covering fraction) at beat maximum which
is consistent with the behaviour observed in the hardness ratios 
(hardening at  maximum). On the other hand, the fits do not indicate 
variations of the  temperature of the black--body component. 
The normalizations of the black--body and of the optically thin component 
 increase at beat maximum and thus the ratio of their bolometric 
luminosities remains constant. This implies that the spectral
changes are due to variations of the absorber and  of  
 the normalizations
of the black--body and of the hard X-ray component. 
The EW of the 6.4\,keV iron line obtained for these two phases does not
vary within errors. The unabsorbed bolometric flux
of the hot
black--body
component ranges between $\rm 1.1-0.64\times 
10^{-10}\,erg\,cm^{-2}\,s^{-1}$ at beat maximum and minimum 
respectively. The lack of clear modulation  at the synodic frequency  in 
the softest band might then be due to the concurrence of phasing of 
the high density absorbing material which increases when
the black--body  area is largest.\\

\noindent RX\,J0558 shows a different   behaviour; 
  the partial
covering absorber (both column density and covering fraction) is 
constant within errors at spin maximum 
and minimum.  The  spin variability is then only   
due to changes in the normalization of  the optically thin component
which varies by a factor of 1.3  between the two 
phases. The black--body
component is constant due to the lack of variability in the LECS data.
Its average unabsorbed bolometric flux is $\rm \sim4\times
10^{-11}\,erg\,cm^{-2}\,s^{-1}$. 
The ratio between this and that of the 
optically thin component slightly increases at spin minimum, due to the 
decrease in the normalization of the hard X-ray emission at this phase.
The EW of the iron $K_{\alpha}$ line does not
change within the errors, although taking the EW and R at their face
values at
spin minimum, a decrease 
might be suggested. The lack of spectral
changes does not allow us to isolate
the secondary pole in the hard X-rays at the minimum of the spin pulse.\\

\noindent In RE\,0751, we infer a different spectral
behaviour
with respect to the other two sources.
 The  temperature of the
black--body component as well as  the EW of the iron 
fluorescent line and reflection do not change with spin phase. The
normalization of the  black--body component is about the same during the
maximum and the dip 
and larger than that during minimum by a factor of about 3. The
unabsorbed bolometric  black--body  flux at 
maximum and dip is $\sim 3\times 10^{-11}\,\rm erg\,cm^{-2}\,s^{-1}$. 
This indicates  that the overall soft spin modulation is essentially due
to changes in the
projected area of the black--body, with no indication of an
additional component. 
The partial absorber largely covers the X-ray source during 
the dip but with a lower column density, suggesting
photo-electric  absorption by an intervening phase--dependent and very
localized material. 
The reflection is unrealistically high during the dip, but it indicates 
that the cold material is facing the observer. 
The partial covering absorber is also larger 
(both column density and covering fraction) at spin maximum. 
The emission measure of the optically thin component 
EM=2.5$\pm 0.4\,\times10^{55}$\,cm$^{-3}$, linked to its
normalization,
changes by a factor of 1.2 between maximum
and  minimum, but during the dip it is lower by a factor of 2. 
Why  the emission measure of the hard X-ray component is lower during the
dip phases is not understood. 
The ratio of the bolometric fluxes of the 
black--body  and the optically thin component is about unity during the
dip and it is much lower during maximum and minimum (0.5 and 0.2
respectively). 
Then, the overall pulsation at higher
energies is due to absorbing  material which is 
different from that causing the dip. The different shapes of the
modulation between 0.4 and 5\,keV are likely due to an even more complex
and structured absorbing material, for which our simple modeling
of the spectra at the three spin phases cannot account.
Indeed Duck et al. (1994), and similarly 
Kiziloglu et al. (1998), use in their spectral fits two columns  
with different  densities ($10^{22}$ and $10^{23}\,\rm cm^{-2}$), but 
the latter is not constrained from their data. 
However, and unlike Duck et al. (1994), who conclude that the 
general rotational modulation is only due to   the normalization of
each 
spectral component, we find that it is true only for the black--body.

\begin{table*}[ht!]     
\centering 
\caption{Spectral fits to the maximum and minimum  spectra 
of RE\,0751, RX\,J0558 and RX\,J1712. For RE\,0751 the dip spectrum fit
is also reported.} 
\begin{tabular}{cccc}
\hline
\hline
& & &\cr  
Parameter & RE\,0751+14  & RX\,J0558+53  & RX\,J1712-24 \cr
          & Maximum ~~~~ Minimum ~~~~ Dip & Maximum ~~~~ Minimum & Maximum 
~~~~ Minimum \cr 
& & &\cr  
\hline
& & &\cr  
$\rm N_{H(pcfabs)}^{1}$ & 9.0$^{+3.3}_{-1.4}$ ~~~~ 8.3$^{+4.8}_{-3.8}$ 
~~~~ 1.0$^{+0.3}_{-0.2}$  & 8.3$^{+5.1}_{-4.9}$ ~~~~ 6.1$^{+2.9}_{-2.0}$      
&  16.3$^{+4.5}_{-3.6}$ ~~~~ 9.6$^{+4.3}_{-3.4}$ \cr
& & &\cr  
 C$_F^{2}$ & 0.51$^{+0.06}_{-0.08}$ ~~~~ 0.36$^{+0.09}_{-0.12}$ ~~~~ 
0.72$^{+0.04}_{-0.03}$  & $0.40^{+0.06}_{-0.11}$ ~~~~ 0.48$^{-0.06}_{+0.08}$ 
& 0.41$^{+0.03}_{-0.02}$ ~~~~ 0.26$^{+0.03}_{-0.02}$\cr
& & &\cr  
$\rm kT_{BB^3}$ & 58$^{+5}_{-6}$ ~~~~ 56$^{+14}_{-12}$ ~~~~ 
53$^{+7}_{-6}$   & 77$^{-11}_{+10}$ ~~~~ 77$^{+8}_{-7}$ &  
105$^{+16}_{-13}$ ~~~~ 108$^{+17}_{-14}$\cr
& & &\cr  
 $\rm L_{BB}/L_{hard}$ & $\sim$0.5 ~~~~ $\sim$0.2 ~~~~ $\sim$1 & $\sim$0.4 
~~~~ 0.6   & $\sim$0.6 ~~~~ $\sim$0.6\cr
& & &\cr  
 EW$^4$  & 175$^{+57}_{-54}$ ~~~~ 165$^{+62}_{-61}$ ~~~~ 
196$^{+142}_{-158}$ & 290$^{+66}_{-80}$ ~~~~ 260$\pm$90 & 
152$^{+50}_{-51}$ ~~~~ 
174$^{+56}_{-29}$ \cr
& & &\cr 
R$^{5}$ & 0.7$^{+0.96}_{-0.7}$ ~~~~ $<0.7$ ~~~~ 7.7$^{+0.1}_{-1.5}$ & $<2$ 
~~~~ 
$<0.7$ & 1. ~~~~ 1. \cr
& & &\cr 
\hline
& & &\cr
$\chi^2_{red}$  & 0.98 ~~~~ 1.20 ~~~ 0.96 & 1.01 ~~~~ 1.23  & 1.11 ~~~~
0.96 \cr
& & &\cr
\hline
\hline
\end{tabular}
\par
\begin{flushleft}
$^1$ Column density of the partial absorber in units of 
10$^{22}\,\rm cm^{-2}$.\par
$^{2}$ Covering fraction of partial absorber.\par
$^3$ Temperature in units of eV.\par
$^4$ Equivalent width of the 6.4\,keV Gaussian in units of eV.\par
$^5$ Relative normalization of the reflection component in units of 
2$\pi$.
\end{flushleft}
\end{table*}

\section{Discussion}

 In these systems, the pulsations are complex and
different from  each other.The basic difference is that  RX\,J0558 
and RE\,0751 are disc--fed systems, while RX\,J1712 is a 
disc--less accretor. All of them are characterized by a soft and relatively 
hot (60-100\,eV) and a hard (13--30\,keV) optically thin  component
whose behaviour at the  dominant  periodicity reveals great
differences. \\

\noindent 
RX\,J0558 was known to show a single-peaked pulsation at the spin period 
of the white dwarf in the hard X-rays, but 
a double--peaked pulse in the soft bands. We cannot derive constraints on
the soft X-ray variability due to limitations of the LECS data.
 The lack of spin variation of the partial absorbing material  
 does not fit with the general picture of IPs, where 
spin pulses are  due to aspect angle changes of the emitting regions 
combined with  photoelectric absorption in the
accretion curtain fed by the disc.  
 Allan et al. (1996) and Norton et al. (1999) proposed that rapid
rotators, like RX\,J0558, possess weak magnetic fields,  allowing the 
accretion material to flow onto large surface areas of the white
dwarf. In this configuration 
the optical depth along the curtain is smaller than perpendicularly and,
hence, absorption effects, as seen by the observer, can be opposite to
those encountered in high field IPs. This modified accretion scenario is
not supported by
either our observations 
or ROSAT data (Haberl et al. 1994), where the 0.1--2.4\,keV
pulses did not show any energy dependence. Also, optical polarized
radiation has been detected in RX\,J0558 (Shakhovskoj \& Kolesnikov 1997),
indicating that the magnetic field can be as high as that in RE\,0751
and RX\,J1712. Furthermore,
since RX\,J0558 shows  a double humped soft X-ray light curve in other
observations, both poles are visible and hence its inclination is
likely larger than
40$^o$. The lack of X-ray eclipses  also gives an upper
limit
of 80$^o$ to the binary inclination. In this range, we should expect
changes in the absorption with the spin phase,
unless the absorbing material is far away from the X-ray emitting
region. For this system the black--body and optically thin 
component mostly balance and hence, we use the hard X-ray luminosity to
estimate the accretion rate. Assuming a white dwarf mass in the range
0.6-1.1$\rm M_{\odot}$ (the latter from Ramsay 2000), we derive $\rm \dot
M \sim 3-5\times 10^{14}\,(d/100pc)^2\,g\,s^{-1}$. 
Due to the lack of magnetic field measures and knowledge of the
distance, we cannot derive the Alfven radius to infer up to what distance
the absorbing material might be located. \\

\noindent   In RE\,0751, the spin modulation is highly 
structured and energy dependent, confirming previous results 
(Duck et al. 1994).  We  find that details of the medium 
energy  pulsation change with respect to previous observations and we 
derive simultaneous information on the soft and hard spectral components. 
The energy-dependent pulses are affected by absorption from
material partially covering the X-ray source, but
with a phase behaviour (minimum at spin
minimum), which is not expected from the classical accretion curtain
scenario. 
The dip feature is due to further intervening
phase--dependent material in the accretion flow which mostly
(70$\%$) covers the X-ray source.
The soft X-ray pulsation is instead due to variations in the 
projected area of the X--ray irradiated polar region. 
A strong magnetic field (9-21\,MG) has been estimated, with  a
dipole offset of 30$^{\rm o}$ and an
inclination angle of 60$^{\rm o}$, allowing
 two extended arc-shaped accreting regions,  
placed ahead of the magnetic poles (Potter et al. 1997).
In this model, material couples at large radii and travels  
out of the orbital  plane, producing large absorption
effects when the accretion region passes through the line of sight at
some phases and in particular the dip.
At pulse maximum and at the dip, the projected area of the black--body
component and the absorption are largest,
in agreement with Potter et al.'s model.
The ratio of the bolometric black--body and the hard X-ray luminosities is 
quite low  and different from 
that derived by Duck et al. (1994) who instead found 2.2-1.5,
compared to our 0.5. We note that RE\,0751 was at a similar flux 
level as that observed during ROSAT observations. The differences
are, however, within the errors of our bolometric fluxes. Given
the lack of balance between these fluxes in our fits, we use  again
the
bolometric luminosity of the hard
X-ray component to estimate the mass accretion rate. Adopting  a distance
of  400\,pc (Patterson 1994),  we derive a luminosity of 
8.3$\times 10^{32}\,\rm erg\,s^{-1}$, and, assuming a white dwarf mass
between 0.6-1.35\,$\rm M_{\odot}$ (the latter from Cropper  et al. 1998),
we
derive $\rm \dot M$ = 2-8$\times10^{15}\,\rm
g\,s^{-1}$. Adopting a value for the magnetic moment of
3$\rm \times10^{33}\,G\,cm^3$, obtained for a 15\,MG field
and an average radius  corresponding to the above mass range, and using
the derived mass accretion
rate, we find that the Alfven radius 
$\rm r_{A}=0.52 (\mu_{WD})^{4/7}\,(2GM_{WD})^{-1/7}\,(\dot M)^{-2/7} 
 \sim 4\times 10^{10}\,cm$.
This implies that indeed the material can be captured at large radii. \\

\noindent In RX\,J1712, the pulsation at the synodic period indicates
 a disc--less accretion, which switches from one pole to the
other each half of a beat cycle. Hence, differently from
Polars or disc--fed IPs, where the spin modulation is
purely geometrical, in stream--fed systems the  poles experience 
changes in mass accretion rate. 
 As already pointed out by Buckley et
al. (1997) and Hellier \& Beardmore (2002), the pulsation is not total,
implying that an accreting region is visible during
the inter--pulse phases.  The absence of an orbital variability
typically expected in a disc--less configuration indicates that 
 RX\,J1712 is a  low inclination system.  Indeed, a 10$\rm ^{o}$
inclination
is reported by Buckley et
al. (1995) and hence  the secondary pole is
never
observed and the primary pole is mostly visible throughout the spin
cycle. Our data do not support an additional
emitting region in both soft and hard X-rays. It is thus possible
that  the pole flipping is not
 total, with the primary pole still accreting but at a lower level, so
that at beat minimum we are still viewing parts of the accretion flow.
 Then,
when the visible pole is accreting at a higher rate (beat maximum), a
higher flux and higher absorption are expected as is observed.
 To account for the low amplitude of the beat modulation, 
 Hellier \& Beardmore
(2002) suggest that the accretion involved in the pole--flipping is 
only 25 percent of the total flow. The remaining
material  is circulating in a diamagnetic blob flow threading the
magnetosphere and forming  a non--coherent accretion curtain, thus making 
 the spin pulsation undetectable. 
If so, it is reasonable to expect  spectral changes 
during the faint phase but  our spectral fits do not show this 
in both soft and hard X-ray components.
 Since the bolometric luminosities of the black--body
and optically thin components roughly balance,  we use the hard
X-ray
bolometric luminosity to 
estimate the accretion rate. We assume a conservative
distance of  50\,pc (V\"ath 1997) and  a 0.7$\rm M_{\odot}$ white dwarf
(Ramsay 2000), thus giving 
$\rm \dot M \sim 3\times 
10^{14}\,g\,s^{-1}$, which changes by a factor of $\sim$ 1.7 between maximum
and minimum of the beat pulse.

\section{Conclusions}

 We have presented the first simultaneous soft and hard X-ray data of the
three
soft X-ray IPs, RE\,J0751, RX\,J0558 and RX\,J1712, over the wide energy 
range of the BeppoSAX satellite.  All three 
sources have strong X-ray modulation at pe\-riods identified as 
the white dwarf rotations in RX\,J0558 and RE\,0751, and as the synodic 
period in RX\,J1712. 
 We have inferred the temperatures of both post--shock
flow and of the irradiated polar region of the white dwarf.
The temperatures of the black--body component are very high, 
 up to 80-100\,eV in RX\,J0558 and RX\,J1712, much higher than those found
in the Polars. The temperature of the hard X-ray emitting region is 
found between 13-30\,keV. Furthermore, a Compton 
reflection component is definitely present in RX\,J0558 and
RE\,0751, while in RX\,J1712 it is favoured. The reflection  likely
originates at the white dwarf surface.

\noindent In the three systems, we find differences in the
periodic behaviour of the spectral components. We find that the absorbing
material partially 
covering the X-ray source is phase--dependent 
in RE\,0751  and RX\,J1712 but not in RX\,J0558.  
The column densities can be as high as 10$^{23}\,\rm cm^{-2}$, as  in 
RX\,J1712, which can heavily hamper the detection of soft X-rays
in other hard X-ray IPs. 

\noindent Although these soft X-ray IPs share common properties, their
X-ray temporal and spectral characteristics  suggest accretion
patterns that cannot be reconciled with a single and simple
configuration.

\begin{acknowledgements}
DdM and GM acknowledge financial support from ASI. 
\end{acknowledgements}

\end{document}